\documentclass[a4paper,12pt]{article}
\usepackage{amsmath,amssymb}
\usepackage{epsfig,graphicx}
\usepackage[german, english]{babel}
\usepackage{ifthen}
\usepackage{a4wide}
\newcounter{fig}   \newcommand{\lbfig}[1]{\refstepcounter{fig}
\label{#1} }
\newcommand{\vphi}{\varphi}
\newcommand{\Tr}{{\rm Tr}}

\newcommand{\bea}{\begin{eqnarray}}
\newcommand{\eea}{\end{eqnarray}}
\newcommand{\be}{\begin{equation}}
\newcommand{\ee}{\end{equation}}

\begin{document}
\begin{titlepage}
\vspace*{2cm}

\begin{center}
{\bf\large Non self-dual and self-dual $SU(2)$ Calorons}
\vspace{2.0cm}

{\sc\large Ya. Shnir}\\[12pt]
{\it  Institut f\"ur Physik, Universit\"at Oldenburg} 
{\it D-26111, Oldenburg, Germany}

\end{center}

\date{~}

\bigskip
\begin{abstract}
New static regular axially symmetric solutions of $SU(2)$
Euclidean Yang-Mills theory are constructed numerically. They
represent calorons having trivial Polyakov loop at spacial
infinity. The solutions are labeled by two integers $m,n$. It is
shown that besides known, charge one self-dual periodic instanton
solution, there are other non-self dual solutions of the
Yang-Mills equations naturally composed out of pseudoparticle
constituents.
\end{abstract}
\bigskip
\end{titlepage}



\section{Introduction}
The interplay between properties of self-dual BPS monopole
solutions  \cite{BPS} and instantons
\cite{Instanton,Witten} caused a lot of attraction over
last decade. It was shown that exact caloron solutions, i.e., the
periodic instantons at finite temperature on $\mathbb{R}^3\times
S^1$, for which component $A_0$ approaches a constant at spacial
infinity \cite{HarrShep78,Gross83}, $A_0 \to 2 \pi i \omega =
2 \pi i  \omega^a \sigma^a$,
are composed out of
Bogomol'nyi-Prasad-Sommerfeld (BPS) monopole constituents
\cite{Baal}. This periodic array of instantons corresponds to the
non-trivial Polyakov loop (holonomy) around $S^1$ at spacial
infinity. In the periodic gauge $A_\mu(\mathbf{r}, x_0+T)=
A_\mu(\mathbf{r}, x_0)$ the Polyakov loop operator is defined as
 \be \label{Polakov-loop}
{\cal P}({\bf r}) = \lim_{r\to\infty} P \exp\left(
\int\limits_{0}^{T} A_0({\bf r},x_0)d x_0\right)\, , \ee where $T$
is the period in the imaginary time direction, which is related
with finite temperature $\Theta$ as $T=1/k\Theta$, and $P$ denotes
the path ordering. Non-trivial value of $\mathcal{P}$ acts like a
Higgs field in adjoint representation labeling the vacua, because
under a gauge transformation $U({\bf r})$ it transforms as \be
{\cal P}({\bf r}) \to U({\bf r}) {\cal P}({\bf r}) U^{-1}({\bf r})
\ee Alternatively, one can formulate the model in  $\mathbb{R}^4$
by fixing periodicity modulo gauge transformations. Indeed, on the
spacial infinity the temporal component $A_0 =  2 \pi i \omega^a
\sigma^a$
 can be gauged away by a non-periodic gauge transformation
$U({\bf r},x_0)= \exp \{ 2\pi i x_0 \omega^a \sigma^a\}$ and then
\be
A_\mu(\mathbf{r}, x_0+T)= e^{2\pi i  \omega^a \sigma^a} A_\mu(\mathbf{r}, x_0)
e^{-2\pi i \omega^a \sigma^a}
\ee

For the self-dual caloron solutions, considered in \cite{Baal} (so
called KvBLL calorons), the field strength vanishes at spatial
infinity, or, equally, it is a pure gauge there. In this case the
constituents are just BPS monopole-antimonopole pairs. The
property of self-duality allows us to apply very powerful
formalism of ADHM-Nahm construction \cite{ADHM} to obtain
different exact multi-caloron configurations \cite{Baal,BaalFalk} and
analyse properties of the BPS monopole constituents. In
particular, it was shown that, as the size of charge one $SU(2)$
caloron is getting larger than the period $T$, the caloron is
splitting into constituents, which represent monopole-antimonopole
pair configuration. The properties of these saddle point solutions
in related $SU(2)$ Yang-Mills-Higgs (YMH) model were discussed
first by Taubes \cite{Taubes81}, different monopole-antimonopole
YMH systems were constructed numerically in
\cite{Rueber,mapKK,KKS}, both in the BPS limit and beyond.

However, besides the self-dual instantons, also solutions to the
second order Euler-Lagrange equations of the euclidean Yang-Mills
(YM) theory are known \cite{non-sd-Instanton}. Also recently the
non-self dual instanton-antiinstanton pair static configuration
was constructed \cite{TR06}, which represent a saddle point
configuration, the deformation of the topologically trivial
sector.

For the non-self dual instantons the action is finite but the field
strength behaves different at spatial infinity and the action is not
proportional to the Chern-Pontryagin topological charge.

In the present work we study static axially symmetric $SU(2)$ YM
caloron solutions on $\mathbb{R}^3\times S^1$ with trivial holonomy, and
find regular numerical solutions, which are labeled by two integers
$n,m$, as their counterparts in the YMH system, the
monopole-antimonopole chains and the circular vortices \cite{KKS}.
Similar to the case of the axially symmetric instantons discussed in
\cite{TR06}, only $m=1$ solutions are self dual, the calorons
labeled by $m \ge 2$ however are non-self dual. The latter
configurations are composed of constituents and correspond to the
monopole-antimonopole chains and/or to the vortex-like solutions.

In section II we present the action of the euclidean YM theory, the axially symmetric ansatz
and the boundary conditions imposed to get regular solution. We will make a detailed numerical
study of the solutions of the corresponding second order field equations.
In section III we discuss the properties of the caloron solutions.

\section{Euclidean $SU(2)$ action and axially symmetric ansatz}
We consider the usual $SU(2)$ YM action
\begin{equation} \label{S}
S = \frac{1}{2}  \int d^4 x \Tr\left( F_{\mu\nu} F_{\mu\nu}\right)
= \frac{1}{4}  \int d^4 x \left(  F_{\mu\nu} \pm {\widetilde  F}_{\mu\nu}\right)^2
\mp \frac{1}{2}  \int d^4 x \Tr\left(  F_{\mu\nu}{\widetilde  F}_{\mu\nu}\right)
\end{equation}
in Euclidean space $R^3 \times S^1$ with one periodic dimension $x_0 \in [0,T]$ and
in normalization where the gauge coupling $e^2=1$. Here
$su(2)$ gauge potential is $A_\mu = A_\mu^a \tau^a/2$ and the field strength tensor is
$
F_{\mu\nu} = \partial_\mu A_\nu - \partial_\nu A_\mu + i[A_\mu, A_\nu]$.
The topological charge is defined as
\be
Q= \frac{1}{32\pi^2} \varepsilon_{\mu\nu\rho\sigma} \int d^4x
\Tr  F_{\mu\nu} F_{\rho\sigma}
\ee
and for the self-dual configurations  $S= 8 \pi^2 Q$.

To construct new regular caloron solutions of the corresponding
\emph{second order} field equations and investigate dependence of
these solutions on the boundary conditions, we employ the by-now
familiar axially symmetric ansatz for the gauge field
\begin{eqnarray} \label{ansatz}
A_\mu dx^\mu
 =
\left( \frac{K_1}{r} dr + (1-K_2)d\theta\right)\frac{\tau_\varphi^{(n)}}{2e}
&-&n \sin\theta \left( K_3\frac{\tau_r^{(n,m)}}{2e}
                     +(1-K_4)\frac{\tau_\theta^{(n,m)}}{2e}
\right) d\varphi;
\nonumber \\
A_0 = A_0^a\frac{\tau^a}{2}
& = &
\left(K_5\frac{\tau_r^{(n,m)}}{2}+ K_6\frac{\tau_\theta^{(n,m)}}{2} \right) \  ,
\nonumber
\end{eqnarray}
which was previously applied to the Yang-Mills-Higgs system
\cite{KKS}. The ansatz is written in the basis of $su(2)$ matrices
$\tau_r^{(n,m)},\tau_\theta^{(n,m)} $ and $\tau_\vphi^{(n)}$ which
are defined as the dot product of the Cartesian vector of Pauli
matrices $\vec \tau $ and the spacial unit vectors
\begin{eqnarray}
{\hat e}_r^{(n,m)} & = & \left(
\sin(m\theta) \cos(n\vphi), \sin(m\theta)\sin(n\vphi), \cos(m\theta)
\right)\ , \nonumber \\
{\hat e}_\theta^{(n,m)} & = & \left(
\cos(m\theta) \cos(n\vphi), \cos(m\theta)\sin(n\vphi), -\sin(m\theta)
\right)\ , \nonumber \\
{\hat e}_\vphi^{(n)} & = & \left( -\sin(n\vphi), \cos(n\vphi), 0 \right)\ ,
\label{unit_e}
\end{eqnarray}
respectively. The gauge field functions $K_i$, $i=1,\dots,6$
depend on the coordinates $r$ and $\theta$.

Recall that although the ansatz \eqref{ansatz} is static, there is
a time dependent gauge transformation which can eliminate the
temporal component $A_0$ at spatial infinity, then the fields
$A_k$ will have a periodic time dependence modulo gauge
transformation.

Substitution of the axially symmetric ansatz \eqref{ansatz} into definition of the
topological charge $Q$ yields similar to \cite{KKS,TR06}
$$
Q = \frac{n}{2} \left[1-(-1)^m\right]\, ,
$$
that is, the configurations labeled by an even integer $m$,
correspond to the topologically trivial sector and represent
saddle point solutions.

The number of the structure functions of the ansatz \eqref{ansatz}
evidently exceeds what one needs to solve first order self-duality
equations, in components there are only 3 equations on 6
functions,  so the system is overdetermined. Thus, a self-dual
configuration corresponds to reduction of the ansatz
\eqref{ansatz}. Actually, the Harrington-Shepard solution
\cite{HarrShep78} as well as KvBLL calorons \cite{Baal}, were
constructed on the Corrigan-Fairlie-'t Hooft, or Jackiw-Nohl-Rebbi
ansatz \cite{CorrFair}
$$
A_\mu = i \bar \eta_{\mu\nu} \partial_\nu
\ln \phi
$$
and its generalizations. Here the periodic function $\phi$ is a
solution of the Laplace equation as required by self-duality, 
for example, the charge-1 Harrington-Shepard caloron is generated by harmonic fanction
$$
\phi = 1 + \frac{\lambda^2}{2r}\frac {\sinh (2 \pi r/T)}{\cosh (2
\pi r/T)- \cos (2 \pi x_0/T)}
$$
where the constant $\lambda$ depends on the size of pseudoparticle
and on the period $T$.

To satisfy the condition of finiteness of the total Euclidean
action \eqref{S}, we require that the field strength vanishes at
the spatial boundary as $\Tr (F_{\mu\nu} F_{\mu\nu}) \to
O(r^{-4})$ as $r \to \infty$. In the regular gauge the value of
the component of the gauge potential $A_0$ at spatial infinity
approaches a constant, i.e., \be A_0 \to \frac{i
\beta}{2}\tau_r^{(n,m)} \ee This corresponds to the holonomy
operator \eqref{Polakov-loop} \be \label{Polakov-loop-1} \Tr~ {\cal
P}({\bf r}) = \Tr~ \exp\left(\frac{i \beta
T}{2}\tau_r^{(n,m)}\right) = \Tr~ U \exp \left( \frac{i\beta
T}{2}\tau_z \right) U^{-1} = \cos \frac{\beta T}{2} \, , \ee where
$U \in SU(2)$ and $\beta \in [0; 2\pi/T]$. Using the classical
scale invariance we can fix $\beta =1$.

Let us consider deformations of the topologically trivial sector
and the deformations of the caloron solution with trivial holonomy
at infinity \cite{HarrShep78,Chakrabati87}. The latter is defined
as a time-periodic array of instantons directed along the
Euclidean time axis. If this array is infinite, the topological
charge one spherically symmetric Harrington-Shepard caloron
coincides with BPS monopole up to a time-dependent gauge
transformation \cite{Rossi}. Generalization of this solution
corresponds to a time-periodic array of instantons of charge $Q$
\cite{Chakrabati87}.

Then we impose the boundary conditions at infinity such that
\begin{equation}
\begin{split}
{\rm for} \ \ {\rm even} \ \ m=2k \ :& \ \
A_0  \longrightarrow  \beta {\hat e}_r^{(n,m)} =  \beta U \tau_z U^\dagger \   , \ \ \
A_k \ \longrightarrow \ i \partial_k U U^\dagger \ ,\nonumber\\
{\rm for} \ \ {\rm odd} \ \ m=2k+1 \ :& \ \ A_0  \longrightarrow
A_{0 \infty}^{(n,1)} = \beta {\hat e}_r^{(n,m)} \   , \ \ \ A_k \
\longrightarrow \ U A_{k \infty}^{(n,1)} U^\dagger +i
\partial_\mu U U^\dagger \ ,\nonumber
\end{split}
\end{equation}
\noindent where $U = \exp\{-i k \theta\tau_\varphi^{(n)}\}$ and
$A_{\mu \infty}^{(n,1)}$ is the self-dual charge $Q=n$ generalized
caloron solution at spatial infinity
\cite{HarrShep78,Chakrabati87}. We will not require, however, that
the gauge field has to be self-dual, i.e., $F_{\mu\nu} \not= \pm
\widetilde{F}_{\mu\nu}$, in general.

In terms of the profile functions of the ansatz \eqref{ansatz} these boundary conditions
read:
\begin{equation}
\begin{split}
K_1 &\longrightarrow 0\ , \quad
K_2 \longrightarrow 1 - m\ , \quad
K_3 \longrightarrow \frac{\cos\theta - \cos(m\theta)}{\sin\theta}
\ \ ({\rm for} \ \ {\rm odd} \ \ m) \ ,\nonumber\\
K_3 & \longrightarrow \frac{1 - \cos(m\theta)}{\sin\theta}
\ \ ({\rm for } \ \ {\rm even} \ \ m) \ , \quad
K_4 \longrightarrow 1- \frac{\sin(m\theta)}{\sin\theta} \ , \quad
K_5\longrightarrow  1 \ , \ \ \ \ K_6 \longrightarrow 0 \ .
\nonumber
\end{split}
\end{equation}

Regularity at the origin requires
\be
\begin{split}
K_1(0,\theta)=0\ , \ \ \ \ K_2(0,\theta)= 1 \ &, \ \ \ \
K_3(0,\theta)=0 \ , \ \ \ \ K_4(0,\theta)=1 \ , \ \ \ \ \nonumber\\
\sin(m\theta) K_5(0,\theta) + \cos(m\theta) K_6(0,\theta) = 0 &\qquad
\left.\partial_r\left[\cos(m\theta) K_5(r,\theta)
              - \sin(m\theta) K_6(r,\theta)\right] \right|_{r=0} = 0 \nonumber.
\end{split}
\ee
Regularity on the $z$-axis, finally, requires
$$
K_1 = K_3 = K_6 =0 \ , \ \ \  \
\partial_\theta K_2 = \partial_\theta K_4 = \partial_\theta K_5 =0 \ ,
$$

\section{Numerical results}

The regular caloron solutions with finite action density and proper asymptotic
behavior can be constructed numerically by imposing these
boundary conditions and solving the resulting system of 6  coupled non-linear
partial differential equation  of second order.
As usually, to obtain regular solutions we have to fix the gauge
condition as $\partial_r A_r + \partial_\theta A_\theta = 0$ (reduced Lorentz gauge),
or $ r \partial_r K_1 - \partial_\theta K_2 = 0 $ \cite{KKS,KKT} and introduce the
compact radial coordinate $x=r/(1+r) \in [0:1]$.The numerical
calculations were performed with the software package FIDISOL
based on the Newton-Raphson iterative procedure \cite{FIDI}.

The simplest class of the solutions corresponds to the $m=1$. It
turns out that, similar to \cite{TR06}, these solutions are
self-dual. We check this conclusion by numerical calculation of
the integrated action density, as well as direct substitution of
the solutions into the first order self-duality equations.
Furthermore, the $m=n=1$ solution is the Harrington-Shepard
spherically symmetric finite temperature solution
\cite{HarrShep78} of unit topological charge. The  $m=1, n\ge 2$
solutions are axially symmetric and the action density
distribution has a shape of a torus.

The $m\ge 2$ configurations satisfy only the second order
Yang-Mills field equations, thus they are not self-dual. Similar
to their counterparts in YMH theory \cite{KKS}, the solutions with
$n=1, m=2,3,4 \dots$ represent chains of time-periodic arrays of
instantons and anti-instantons of unit charge interpolating on the
symmetry axis. A general property of these solutions is that the
corresponding action density possesses $m$ clear maxima on the
axis of symmetry (see Fig \ref{f-2}). Thus, we can distinguish $m$
individual constituents and identify these with non-self dual
chain of periodic instantons. Also, the topological charge density
possesses $m$ local extremes on the $z$ axis, whose locations
coincide with maxima of the action density. The positive and
negative extremes alternate between the locations of the
individual constituents.


\begin{figure}\lbfig{f-2}\begin{center}
  \includegraphics[height=.32\textheight, angle =-90]{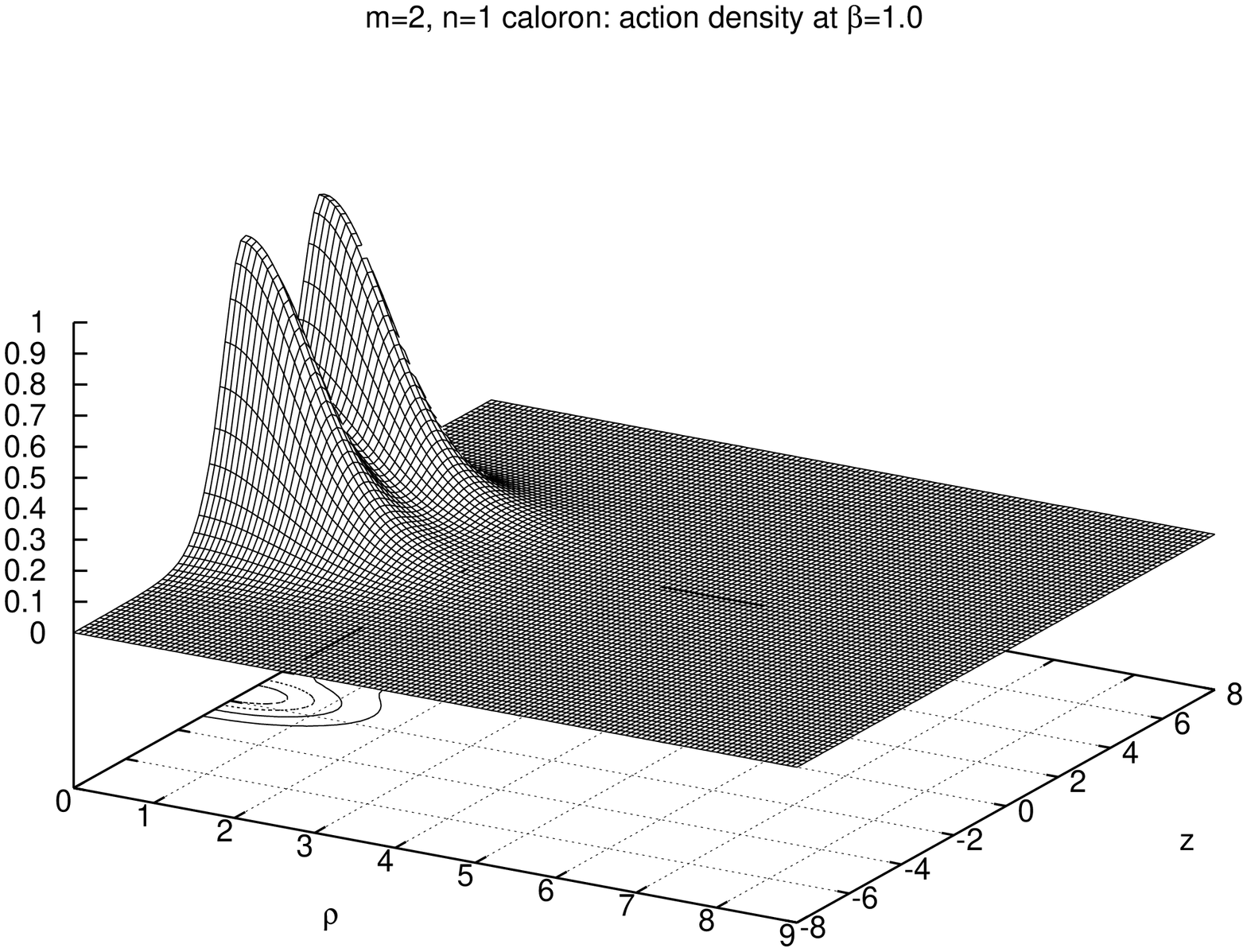}
  \includegraphics[height=.32\textheight, angle =-90]{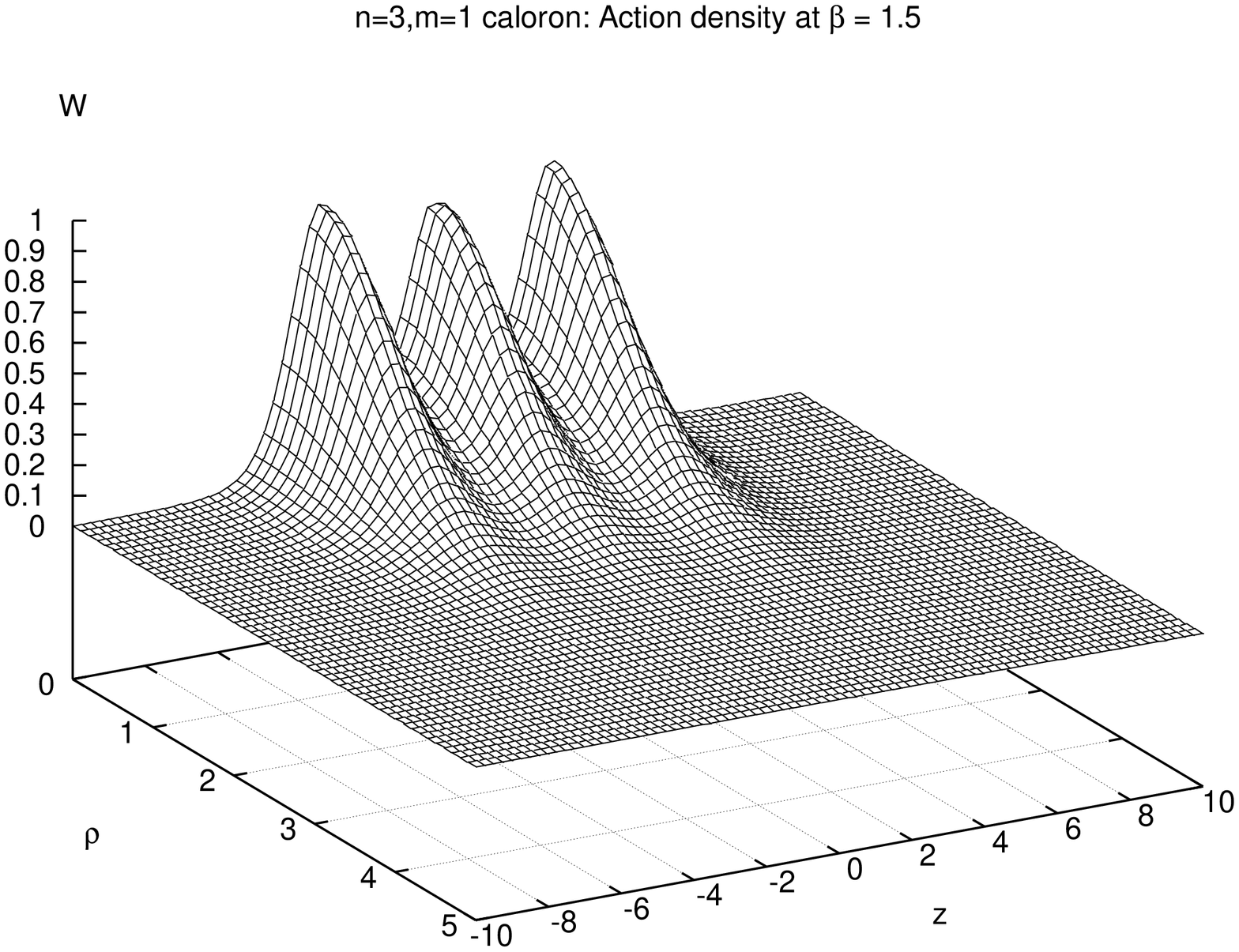}
  \includegraphics[height=.32\textheight, angle =-90]{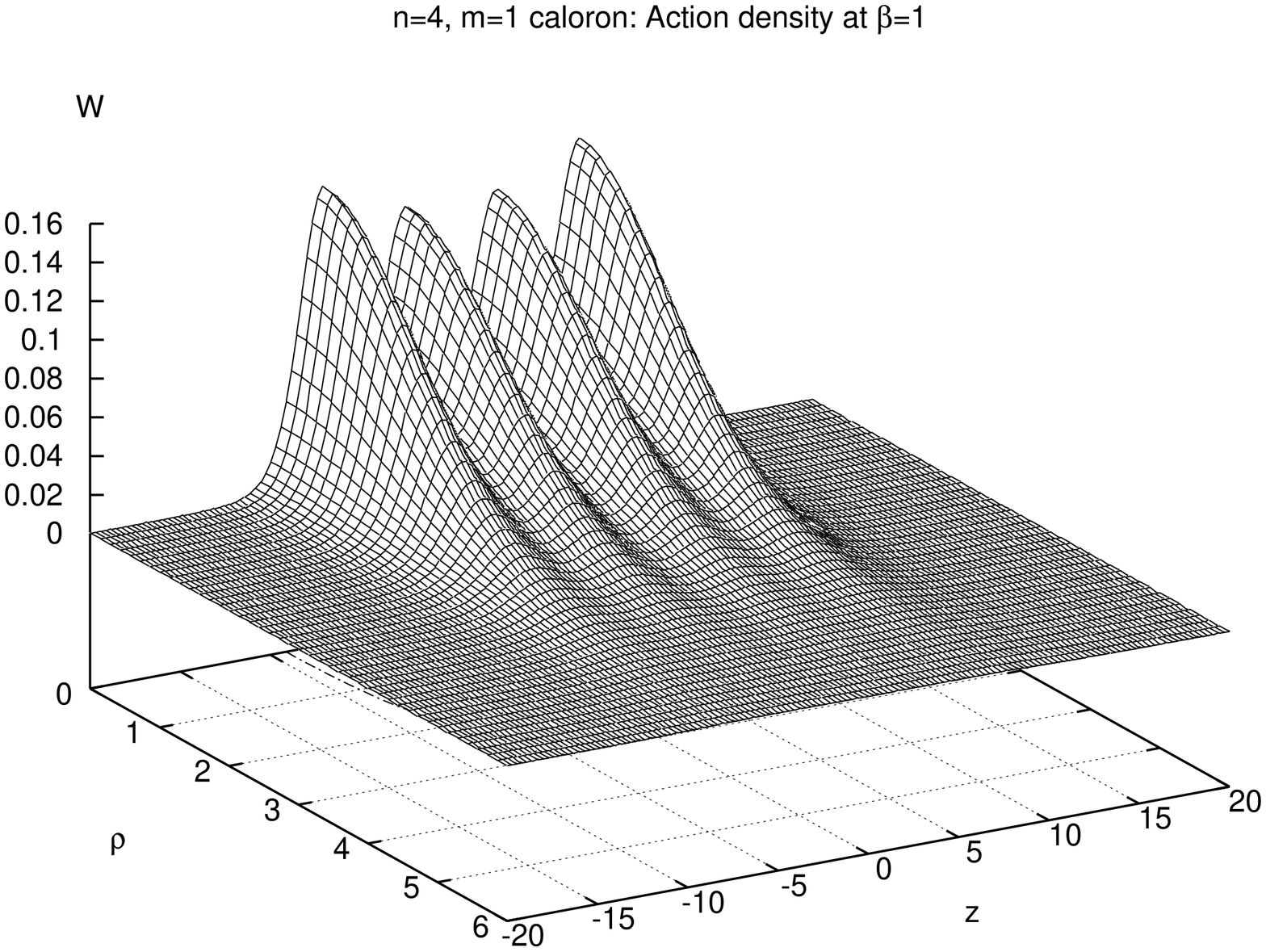}
\end{center}
  \caption{The action
densities of the $m=2,3,4$ and $n=1$ caloron chains are shown in
coordinates $z, \rho$. }
\end{figure}


The same general behavior is observed for all other solutions of
different types. Generally, increasing of the winding number $n$
which is related with topological charge of each individual
constituent pseudoparticle, yields shift of the local extremes of
the action density away from the symmetry axis. For example, for a
configuration with $n=m=3$ (triple charged
instanton-anti-instanton-system) we found three maxima on the $z
\rho$ plane, which corresponds to the surface of triple torus with
one maximum on $x,y$ plane and two other, placed symmetrically
above and below this plane (see Fig \ref{f-4}).  The counterparts
of these configurations in YMH theory are monopole-vortex rings
systems \cite{KKS}. Again, the radius of the tori and relative
distance between their location decreases as $\Delta \rho_0 \sim
1/\beta$.
\begin{figure}\lbfig{f-4}\begin{center}
  \includegraphics[height=.32\textheight, angle =-90]{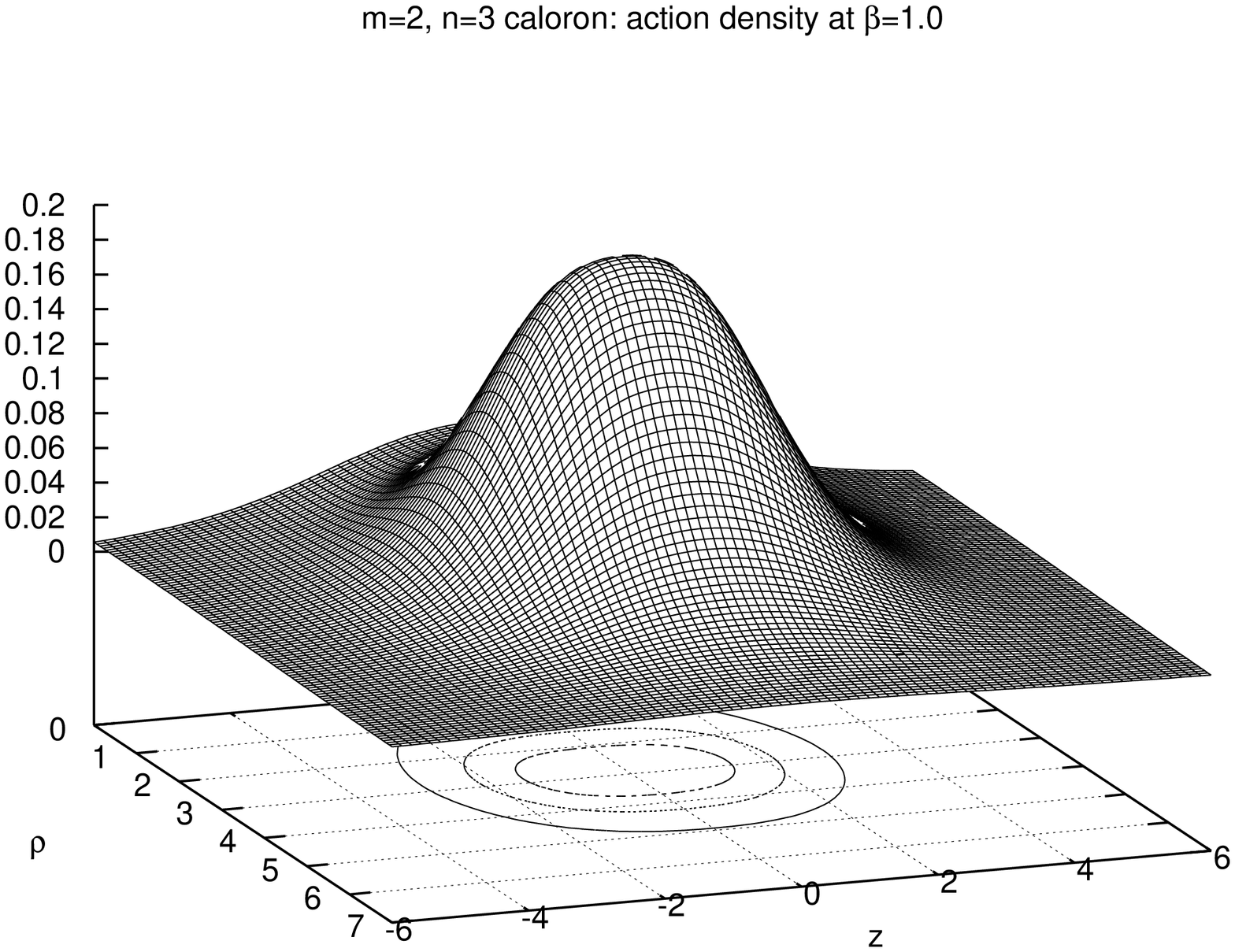}
  \includegraphics[height=.32\textheight, angle =-90]{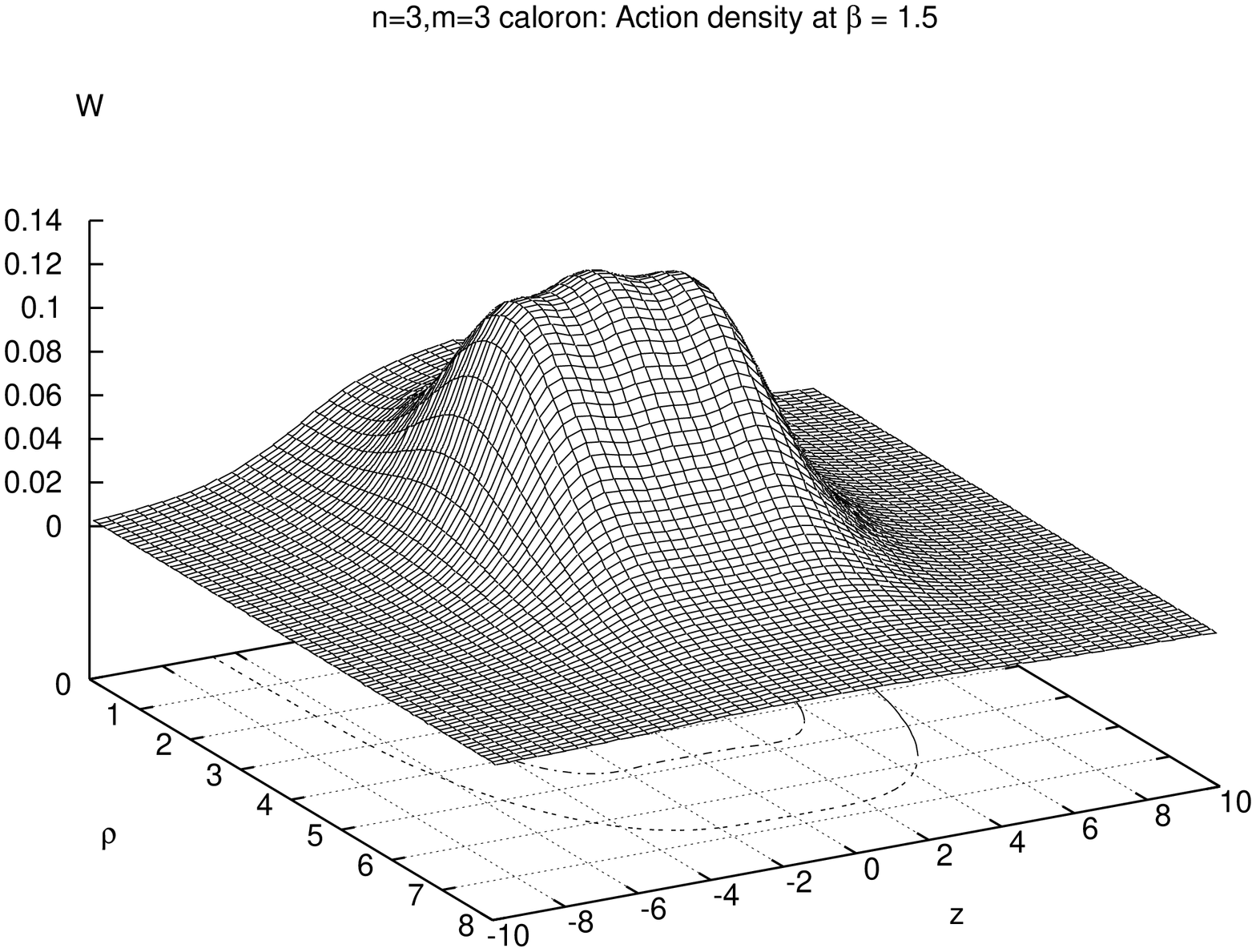}
\end{center}
  \caption{The action density distribution
is plotted for the $n=2$ and $n=3$ calorons with winding number $m=3$,
respectively.
}
\end{figure}
The numerical results indicate that the integrated action of the
$m\ge 2$ configurations for all non-zero values of temperature
remains above the self-duality bound. Since the counterparts of
these solutions in YMH theory correspond to the sphaleron-like
solutions, there is a reason to believe that such caloron
solutions also are unstable, they correspond to saddle points of
the action functional. Note that the variation of the temperature
does not lead to the chain-vortex bifurcations, which were
observed in the YMH systems in external electromagnetic field
\cite{S05} or in the limit of large scalar self-coupling
\cite{KKNS}.

\section{Conclusions}
To summarize, we have constructed axially symmetric caloron
solutions to the d=4 Euclidean $SU(2)$ YM theory by numerical
solution of the second order Yang-Mills equations. Similar to the
monopole-antimonopole axially symmetric solutions to the YMH
theory, the calorons are labeled by two winding numbers $(n,m)$
and the topological charge of the configuration is $Q =
\frac{n}{2} \left[1-(-1)^m\right]$. The action density of the
configuration has non-trivial shape and position of the maxima of
the action functional allow us to identify location of each
individual constituent. Besides configurations with $m=1$, which
are self-dual, the solutions do not saturate the self-duality
bound.

For the chain solutions with $n=1,2$ there are  periodic arrays of
the instantons and the anti-instantons, which are located on the
axis of symmetry in alternating order. For configurations of
higher topological charge the action density forms a torus-like
shape.


The caloron solutions described here are restricted, because the
ansatz \eqref{ansatz} possesses the reflection $Z_2$-symmetry with
respect to the $xy$ plane. This is not the symmetry of the  KvBLL
solution, the latter has only the $O(2)$ symmetry with respect to
the rotation about the axis of symmetry \cite{Baal}. To describe
general non-self dual axially symmetric caloron solutions, also
with non-trivial holonomy, one has to implement an extended ansatz
for the gauge field which includes complete set of 12 profile
functions and consider a different set of the boundary conditions.
The results of the related calculations will be reported elsewhere.

Although both the configurations considered above, and the KvBLL
calorons admit the constituent interpretation with lumps being
associated with monopoles, there is an important difference. The
former caloron solutions, in a general case are defined along the
same positive simple root, which corresponds to a given $SU(2)$
subgroup of $SU(N)$. For example, the configuration with winding
numbers $n=1, m=2$ corresponds to the monopole-antimonopole pair
solution described in \cite{mapKK,KKS}. The monopole constituents
of the $SU(N)$  KvBLL calorons \cite{Baal} are defined in a
different way, e.g., the $SU(2)$ caloron configuration describes
monopole of positive charge embedded along positive simple root
with the asymptotic $A_0 \to \beta$, and a Weyl-reflected
antimonopole with asymptotic $A_0 \to 2\pi /T - \beta$.

It would be interesting to see how the non-self-dual caloron solutions presented here
can be relevant for QCD, in particular how the corresponding saddle point configurations
may contribute to the process of the confinement-deconfinement phase transition.\\
\paragraph{Acknowledgements} 
I am grateful to Pierre van Baal, Falk Bruckmann,
Michael Ilgenfritz, Michael Mueller-Preussker, Werner Nahm, Eugen
Radu and Tigran Tchrakian for useful discussions and comments. I
would like to acknowledge the hospitality at the Department of
Mathematical Physics, NUI Maynooth and Service de Physique
Th\'eorique, CEA-Saclay where this work was begun.

\end{document}